Unraveling the Skillsets of Data Scientists: Text Mining Analysis of Dutch University Master Programs in Data Science and Artificial Intelligence

Short title:

Data science and AI skills of Dutch graduates

Mathijs J. Mol[1], Barbara Belfi[2], Zsuzsa Bakk[3]

Affiliations

1. Amsterdam University Medical Center
2. Leiden University
3. Maastricht University

Corresponding author

Email:z.bakk@fsw.leidenuniv.nl

Contributors:

MJM performed the data curation and data analysis. BB updated the final revision of the ms and added theoretical contribution about educational science. ZB developed the research question, supervised data collection and analysis and contributed to writing the ms.

Acknowledgments:

We would like to thank Adham Kahlawi for insightful feedback about the data analysis and data quality on an initial draft of the ms.

# Abstract

The growing demand for data scientists in the global labor market and the Netherlands has led to a rise in data science and artificial intelligence (AI) master programs offered by universities. However, there is still a lack of clarity regarding the specific skillsets of data scientists. This study aims to address this issue by employing Correlated Topic Modeling (CTM) to analyse the content of 41 master programs offered by seven Dutch universities. We assess the differences and similarities in the core skills taught by these programs, determine the subject-specific and general nature of the skills, and provide a comparison between the different types of universities offering these programs. Our findings reveal that research, data processing, statistics and ethics are the predominant skills taught in Dutch data science and AI master programs, with general universities emphasizing research skills and technical universities focusing more on IT and electronic skills. This study contributes to a better understanding of the diverse skillsets of data scientists, which is essential for employers, universities, and prospective students.

# 1. Introduction

Data science is an interdisciplinary field that leverages computational methods and techniques to derive actionable insights from large, complex datasets (Blei & Smyth, 2017). The rise of big data and the growing use of artificial intelligence (AI) have created an enormous need for professionals with advanced data science abilities (Davenport, 2020). Despite the increasing significance of data scientists in today's society, there is still a noticeable lack of clarity regarding the exact skills and expertise these professionals posses (Lee & Delaney, 2022; Verma, Lamsal & Verma, 2022). One reason for this is that the master programs that train students to become data scientists differ greatly in curriculum and focus (Cho, 2022; Feenstra et al., 2018; Lee & Delaney, 2022). For example, the skillsets gained from a biomedical data science program may vastly differ from those acquired in a marketing analytics program. At the same time Sigelman et al. (2019) state that the general demand of data scientists on the job market has increased by 663% from 2013 to 2018 and the demand of marketing data analysts has increased with 194% in the same timeframe.

Understanding the skillsets that data scientists acquire during their education is essential for multiple reasons. First, universities and other educational institutions can utilize this information to refine and update their curricula, ensuring they remain aligned with current industry needs and expectations. Second, businesses and organizations seeking to recruit data scientists can better assess the qualifications of candidates based on their understanding of the typical skillsets possessed by graduates of these programs. Finally, current and aspiring data scientists can use these findings to identify any gaps in their own skillsets and guide their professional development efforts.

As stated before, to the best of our knowledge, as of now, there is no universally accepted, centralized definition of skillsets or knowledge map for the field of data science or artificial intelligence. A few recent articles have attempted to define the skillsets required in data science and AI, but these efforts remain scattered and inconsistent. One such example is the Initiative for Analytics and Data Science Standards (IADSS), which has laid the groundwork for a knowledge framework for the discipline (Fayyed & Hamutcu, 2020). The IADSS framework theoretically identifies key areas of expertise, such as data engineering, data analysis, data visualization, and machine learning, and their corresponding skillsets. This initiative aims to standardize the understanding of skills required for data science and AI professionals and provides a foundation for future curriculum development and workforce training. In contrast, other authors have taken a more data-driven approach to defining the skillsets needed in data science and AI, by analysing job descriptions and identifying patterns and trends (Mauro et al, 2017, Markow et al, 2017). This approach offers valuable insights into the real-world demands and expectations of employers in the field, making it more closely aligned with current industry needs. These studies have highlighted the importance of not only technical skills, such as programming languages, machine learning frameworks, and big data tools, but also soft skills, such as communication, teamwork, and problem-solving abilities. Despite these efforts, the lack of a unified and comprehensive definition of skillsets in data science and AI remains a challenge.

The primary objective of the present study is to gain more clarity in the skillsets of Dutch data scientists by analysing the content of Dutch university master programs in data science and artificial intelligence. We chose to focus on these study programs due to their explicit commitment to providing students with advanced data science skills. Our investigation employs text-mining techniques, specifically Correlated Topic Modeling (CTM; Blei & Lafferty, 2007), to systematically evaluate the course offerings and curriculum content within these programs from the university webpages. While similar efforts have been made to derive skillsets from US based educational programs (Tang & Sae-Lim, 2016; West 2017) no such efforts have been made for the Dutch context.

As such, our research aims to address the research questions: (1) What skillsets are encompassed within data science and AI master programs offered by Dutch universities? (2) What are the overall differences and similarities in the skillsets learned across different universities? (3) How do the skillsets learned at more general data science and AI programs differ from those learned at more subject-specific data science master programs?

The remainder of this paper is organized as follows. In order to situate our study within the existing academic literature, we first review prior research on data science education and skill development. Next, we provide an overview of the text mining technique we employ, Correlated Topic Modeling (CTM), and discuss its potential for revealing insights into the skills taught in data science and artificial intelligence study programs. Subsequently, we present the results. Finally, we will summarize our findings and discuss the implications for various stakeholders, including universities, businesses, and data science practitioners.

## 2. Literature review

### 2.1. Deriving data science skillsets from educational programs

As a consequence of the ever growing interest in big data and analytics the number of educational programs offering data driven skills has grown immensely in the last 10-15 years globally (Davenport, 2020). This growth has mostly been driven by market needs, and is not preceded by a theoretical development of the field of data science (or as a matter of fact related fields such as business analytics, big data), but rather the definition of the field is driven by the market needs and technology forecasts (for example Gorman & Kimberg, 2014). Consequently, there is a wide variety in the skillsets taught across different data science programs (Tang & Sae-Lim, 2016; West, 2017).

To our knowledge, at present only two studies have investigated the skillsets of data scientists by exploring the content of educational programs through text mining (Tang & Sae-Lim, 2016; West 2017). First, Tang and Sae-Lim (2016) studied the content and structure of data science programs in higher education institutions in the United States. They conducted an exploratory content analysis of program descriptions, curriculum structures, and course focuses of 30 undergraduate and graduate programs in data science from various US based institutions. The study concluded that while the investigated data science programs varied greatly in terms of specific technical skills, mathematical/statistical foundations, and domain-specific knowledge. However, core technical skills, such as programming, data management, and machine learning, were present in most programs. Finally, in all programs, only little emphasize on soft skills such as communication, collaboration, and ethical considerations was found.

Furthermore, West (2017) studied the content of academic curricula from undergraduate and postgraduate programs in data science and big data that was presented on the websites of 320 universities worldwide. Although they also observed a wide variety in the content of the different curricula, they came to the conclusion that core skills such as computer coding, statistics machine learning, visualisation and application innovation were core terms in most study programs.

Finally, Cegelieski and Jones-Farmer (2016) employed a somewhat different approach to investigate what knowledge, skills and abilities should be taught in colleges of business to prepare students for entry-level careers in business analytics. To gain more insight in this, they studied the content of 23 US based undergraduate business analytics programs, employing a survey among 27 experts in data analytics, and content analysis of 215 job postings, requiring a 3 year college degree in business analytics and less than 3 years of professional experience. They concluded that while the findings of the three different study approaches differed slightly, there was a general agreement on the importance of business, analytical and technical skills for the entry-level careers in business analytics.

**2.2. Towards a definition of the skillsets of data scientists**

While there is still much ambiguity regarding the core skills that should be trained in data analytics study programs, interestingly enough, the last decade has seen a development of different organizations and agencies grown out of the need to define and regulate the understanding of knowledge, skills and abilities related to data driven jobs (Davenport, 2020). Such organizations are for example the Institute for Operations Research and management Science (INFORMS) Analytics Certification Job Task Analysis Working Group (Cegielski & & Jones-Farmer, 2016, Davenport, 2020) or the IADSS (Fayyed & Hamutcu, 2020) (Demchenko et al, 2016), the Edison project (Fayyed & Hamutcu, 2020). Also new journals have been started to propagate the development of the field of data science/ data driven thinking, for example the Harvard data science review (started in 2019) or the Journal of big data (that received its first impact factor rating in 2022). Yet due to the rapid development of the field, society-level classifications accepted generally are not yet available (Davenport 2020) and most of the available classifications are US based.

Fayyad and Hamutcu (2020) offer a comprehensive review of current efforts to define the field of data science. They argue that arriving at a definitive description is premature due to the lack of consensus on its constitutive elements. However, they note that most definitions share a common emphasis on the integration of statistical and engineering skills. In the context of curriculum development, the authors reference seminal contributions by Song and Zhu (2016) and De Veaux et al. (2017). Song and Zhu (2016) propose a framework for data science education in the United States, emphasizing four core components: big data infrastructure, big data analytics lifecycle, data management skills, and behavioural disciplines. Conversely, De Veaux et al. (2017) identify six critical undergraduate data science competencies: computational and statistical thinking, mathematical foundations, model building and assessment, algorithms and software foundations, data curation, and knowledge transfer.

Furthermore, the Edison project, an interdisciplinary initiative, aims to establish a comprehensive body of knowledge that brings together essential concepts and principles from computer science, software engineering, information and communication technology (ICT), and related fields (Fayyad & Hamutcu (2020). This collaborative effort seeks to promote a better understanding of the interconnections among these domains, enhancing the overall effectiveness and efficiency of professionals working in these areas. The Edison project primarily focuses on business analytics but also accommodates other domains by reserving placeholders for future integration.

Finally, in their 2018 study, De Mauro and colleagues offer a comprehensive review of literature that aims to define the field of data science. They emphasize a range of essential competencies, including proficiency in big data tools, coding, statistical analysis, research, and data hacking. Additionally, they underscore the importance of ethical management and strategic thinking in data science.

In short, there is a wide range of definitions, attempts on creating taxonomies available, that is specific to any scientific field that is not yet defined/ delimited. As also Fayyad and Hamutcu (2020) highlight, common to all is a combination of statistical and engineering skills. This variety of definitions is specifically relevant from the perspective of this article, as our aim is to understand the skillset of graduates of data science and AI programs that are coming from very different universities ranging from technical universities, to general research universities, but also specific programs in business schools or life science faculties. Given the diversity in literature in defining core and domain specific fields taking a look on program curricula will help us seeing how this diversity is reflected in the Dutch educational system. We aim to explore whether some skills, knowledge and competencies are common across graduate programs, as such constituting a core skillset and the diversity of domain specific knowledge taught.

## 3 Methods

In order to answer the research questions, information is gathered about the content of data science and artificial intelligence master programs throughout the Netherlands. This information is gathered in the form of descriptions of programs and courses. Thereafter, text analysis is performed using corelated topic modeling (CTM). Lastly, the posterior assignments obtained from the analysis have been used to highlight the inter-university differences. All the data and code for the analysis is openly available on OSF: https://doi.org/10.17605/OSF.IO/96MQ3.

The CTM is a hierarchical model of document collections that is developed from the Latent Dirichlet Allocation (LDA) based topic model (Blei et al. 2003) by allowing the topics to be correlated. LDA is a generative probabilistic model of clustering unstructured data. In this model, documents are represented as random mixtures over latent topics, where each topic is characterized by a distribution over words. Using such a latent variable based approach allows for a probabilistic classification of documents (in our case course and program descriptions) over multiple topics. The CTM models the words of each document from a mixture model. The components of this mixture model are shared by all documents in the collection, therefore the mixture proportions are document specific random variables. The CTM allows for multiple topics with different proportions for each document that are allowed to correlate. Thus, it allows to capture the heterogeneity in grouped data that show multiple latent patterns.

## 3.1 Data collection

The total dataset consisting of 1009 course and program descriptions that correspond to 41 master study programs have been manually collected from university websites. These universities all are part of 'De Vereniging van Universiteiten' (VSNU), which is a trade group of ten government-funded universities, three special universities and an open university. We chose to only use programs of universities that are part of this association for reliability reasons. In Table 1 a list is shown of the universities together with the number and type of course and program description per university. For one university, no type of course was stated at the official university website. Therefore, a column has been added to the table containing course descriptions that could either be core courses or electives.

**Table 1**

*Number of program and course descriptions per university included in the analysis*

|  | Program | Core course | Elective | Core or elective |
|---|---|---|---|---|
| Delft University of Technology (TU Delft) | 1 | 10 | 66 | 0 |
| Eindhoven University of Technology (Tu/e) | 5 | 32 | 45 | 0 |
| Leiden University (LU) | 3 | 33 | 68 | 0 |
| Maastricht University (MU) | 5 | 26 | 76 | 0 |
| Radboud University Nijmegen (RAD) | 2 | 5 | 13 | 45 |
| Tilburg University (Til) | 5 | 39 | 64 | 0 |
| University of Groningen (RUG) | 2 | 22 | 29 | 0 |
| University of Twente (UT) | 6 | 34 | 87 | 0 |
| University van Amsterdam (UVA) | 5 | 49 | 37 | 0 |
| Utrecht University (UU) | 2 | 6 | 49 | 0 |
| Vrije Universiteit (VU) | 4 | 24 | 67 | 0 |
| Combination: Vrije Universiteit, Erasmus University of Rotterdam and Universiteit van Amsterdam[a]   (Combi) | 1 | 20 | 22 | 0 |

*Note.* The number of program descriptions equals the number of programs selected for this study

[a]This is a program offered by three different universities and is thus a combination of institutions

### 3.2 Data processing

After collecting the descriptions, some altercations have been made to the text in order to obtain more optimized results. The specific changes are stated in Table 2. The text analyses is based on a count of words, defined as "Term Frequency". To optimize the results we combined manually words that occur frequently and have a special meaning in the combined form, for example: "statistical_learning" or "artificial_inteligence". Furthermore some plural forms have been changed into singular form. Words that did not hold any information and numbers in general have been removed and all letters have been set to lower case.

**Table 2**

*Manual changes made to the texts as part of data pre-processing*

| Change | Words |
| --- | --- |
| Plural to singular | models, systems, sets, problems, networks, games |
| Words combined | machine learning, deep learning, statistical learning, data science, computer science, artificial intelligence, data mining, text mining, time series, neural network, research project, distributed computing, natural language processing, probabilistic theory, distributed system, critical thinking, decision making, skill sets, ad hoc |
| Words deleted | will, course, courses, student, students, able, university, master, can, skills, work, new, use, used, using, also, different, learn, learning, part, master's, understand, one, two, game, topics, understanding, based, many, several, exam, make, discussed, ad hoc |

# 4 Results

Figure 1 presents the coherence scores for CTM with 2 to 30 clusters. While the optimum is at 13 topics, at 7 topics the graph shows a clear elbow, as such we will investigate both the 7 and 13 cluster solution.

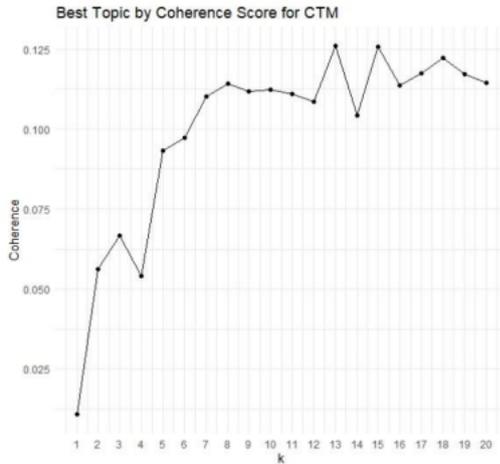

Figure 1. Plot of the coherence scores per number of K topics for the CTM

In Table 3 the results for K = 7 are shown. We see a large number of subdomains and skills sets being described already by the 7 correlated clusters. Apart from the first topic describing core terms, every topic denotes a domain within data science and artificial intelligence, meaning skill sets of this field are being expounded in the descriptions. Interestingly enough the sixth topic is about ethics which is a less often discussed subdomain yet with a growing societal relevance.

When compared to the CTM model where K = 13 (see Appendix C), we see there is a lot of overlap with the CTM model where K = 7. Main differences include ethics not being denoted in the K = 13 model and more subdomains being described in the more complex model such as machine learning, natural language processing, AI/optimization, the human brain, information security and the business domain. Since the CTM model where K = 7 gives a sufficient overview of the corpus, this model has been chosen as the main model.

**Table 3**

*Top ten most common words per topic for the CTM where K = 7 with the topic label in last row*

|   | Topic 1 | Topic 2 | Topic 3 | Topic 4 | Topic 5 | Topic 6 | Topic 7 |
|---|---------|---------|---------|---------|---------|---------|---------|
| 1 | research | model | data | algorithms | data | system | research |
| 2 | project | Data | model | machinelearning | system | artificialintelligence | model |
| 3 | datascience | analysis | techniques | data | information | data | social |
| 4 | business | techniques | process | techniques | software | health | system |
| 5 | data | image | deeplearning | methods | business | concepts | knowledge |
| 6 | programme | statistical | naturallanguageprocessing | model | design | datascience | design |
| 7 | knowledge | theory | language | knowledge | services | ethical | network |
| 8 | scientific | methods | machinelearning | programming | web | decisions | data |
| 9 | development | linear | methods | problem | model | privacy | methods |
| 10 | thesis | computer | datascience | datamining | distributed | problem | human |
| Label | *Core elements of data science* | *Statistics* | *Data processing techniques 3* | *Data processing techniques 4* | *Electronics-IT* | *Ethics* | *Research* |

When we look at the indicated skill sets in Table 3, we can clearly see that there is a wide variety of skill sets needed to complete a data science master program in the Netherlands. It displays the core elements of data science, but also the subdomains of which some knowledge is needed. To clarify this, a distinction has been made between the two and are shown in Table 4.

**Table 4**

Sorted skill sets based on manually named topics according to CTM

| *Data science specific skill sets* | *Skill sets of subdomains* |
|---|---|
| Network (analysis) | Law |
| Deep learning | Health |
| Data mining | Research |
| Natural language processing | Business/marketing |
| Artificial intelligence | Ethics |
| Machine learning | Research |
| Data analysing | |
| IT | |
| Statistics | |
| Data processing techniques | |

Skill sets that correspond to the mentioned fields in Table 4 are taught in data science related master programs throughout the Netherlands. Based on these results, it might be useful for employers to investigate these terms closer to get a better grasp on what data science graduates have to offer, in order to match the skill sets to the concerning vacancies. However, there is quite a variability between universities in the core and subdomain skills they prioritize. In order to have a grasp on this diversity we continue with switching our focus to universities.

We classified the universities based on the most prominent topics in their course and program descriptions. The classification is performed by taking the mean of the posteriors of all descriptions per university per topic averaged across the programs. As such, this profile describes not the whole university, just the investigated programs.

We see in Table 5 that there are a lot of differences in order between the general outcome and the individual universities. Core terms, research and ethics are very prominent topics for multiple universities. A small cluster consisting of Rad, Til, RUG and UU that all share the first two topics, namely research and ethics, is also uncovered. However, apart from these two corresponding topics, the order of the topics that come after differ a lot. Lastly, the two technical universities both score high on electronics/IT, which is the more technical topic.

**Table 5**

*Order of most important manually labelled topic terms per university based on posterior values for the CTM*

|    | University | First | Second | Third | Fourth | Fifth | Sixth | Seventh |
|----|---|---|---|---|---|---|---|---|
| 1  | TU Delft | Electronics/IT | Data processing techniques_4 | Research | Statistics | Ethics | Data processing techniques_3 | Core terms |
| 2  | TU/e | Core terms | Electronics/IT | Research | Data processing techniques_4 | Ethics | Statistics | Data processing techniques_3 |
| 3  | LU | Core terms | Data processing techniques_4 | Research | Statistics | Electronics/IT | Data processing techniques_3 | Ethics |
| 4  | MU | Data processing techniques_4 | Data processing techniques_3 | Ethics | Statistics | Core terms | Research | Electronics/IT |
| 5  | RAD | Research | Ethics | Data processing techniques_4 | Core terms | Data processing techniques_3 | Statistics | Electronics/IT |
| 6  | Til | Research | Ethics | Core terms | Statistics | Electronics/IT | Data processing techniques_3 | Data processing techniques_4 |
| 7  | RUG | Research | Ethics | Core terms | Data processing techniques_3 | Statistics | Data processing techniques_4 | Electronics/IT |
| 8  | UT | Data processing techniques_3 | Electronics/IT | Statistics | Ethics | Research | Core terms | Data processing techniques_4 |
| 9  | UvA | Core terms | Research | Ethics | Data processing techniques_4 | Statistics | Data processing techniques_3 | Electronics/IT |
| 10 | UU | Research | Ethics | Statistics | Data processing techniques_4 | Core terms | Data processing techniques_3 | Electronics/IT |
| 11 | VU | Research | Core terms | Data processing techniques_4 | Statistics | Data processing techniques_3 | Electronics/IT | Ethics |
| 12 | Combi * | Statistics | Research | Core terms | Data processing techniques_4 | Ethics | Electronics/IT | Data processing techniques_3 |

* Combi is defined as one program offered by three universities simultaneously UvA, VU and Erasmus university Rotterdam. The second line represents a manually added key term describing the main topics per cluster

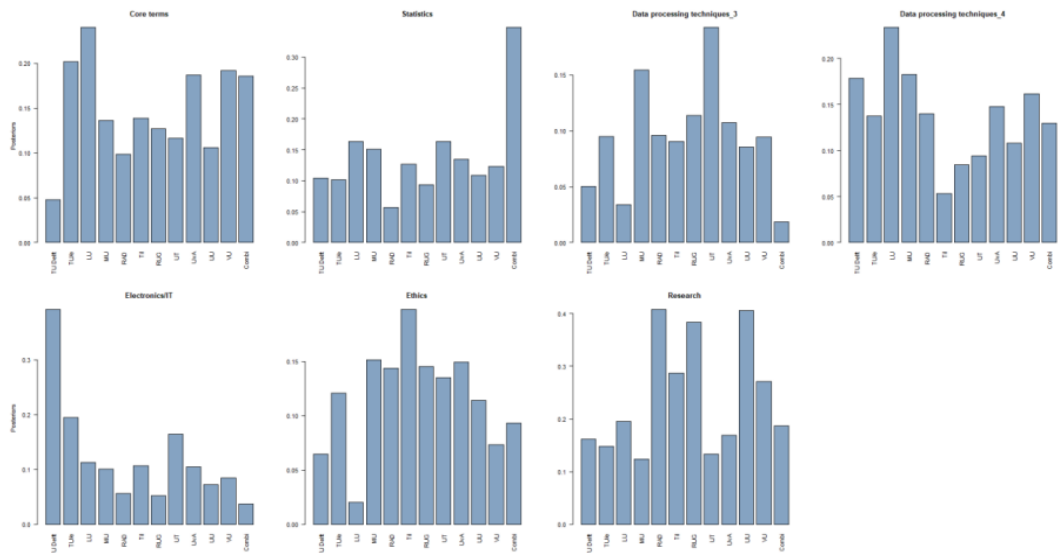

Figure 2. Seven bar plots of the posteriors per university for every manually labelled topic of the CTM outcomes

In Figure 2, we switch the lenses for the presentation to show the prevalence of the seven topics across the universities. These visualisations are plotted to help identify which universities are similar in scoring high on a given topic. Note that when universities score high on a certain topic, it means that their data science and artificial intelligence programs are more oriented towards the described topic. It does not mean that the entire university is oriented towards the concerning domain. The bar chart clearly visualizes the diversity among universities with regard to the prevalence of the different topics.

TU Deft stands out by the relative low prevalence of core terms coupled with the highest prevalence of the Electronics/ IT related topic. On IT/electronics unsurprisingly the second highest score is for Tu/E. The prevalence of the second cluster, focusing on statistics seems to be very balanced across universities, with the Combi program being the only one that focuses more heavily on this pillar. On the research cluster, the more general universities (RAD, RIU, UU) score higher than the technical universities. The differences between the two data processing clusters are also large between universities, with for example TU Delft, TU/e and LU having more focus on the second data processing pillar, while UT on the 1$^{st}$, with MU, Uva, VU, and TiL a more balanced distribution between the pillars. As such, industries with special interest of skills corresponding to certain pillars can be well informed by such a distribution of focus between programs. LU scores specifically low on the Ethics dimension- but we need to mention that for this university only the more technical programs were selected (not including the law faculty).

## 5 Discussion and Conclusion

In this article, we set out to examine the skillsets of data science graduates in the Netherlands. Our goal was to gain a better understanding of what skillsets are taught within data science and AI master programs offered by Dutch universities and how they differ or are similar across institutions. Additionally, we wanted to compare the skillsets learned in more general data science and AI programs with those learned in subject-specific data science master programs. To achieve this, we employed a text mining technique called CTM (Correlated Topic Modeling) on 1009 program and course descriptions from data science and artificial intelligence master programs of 7 Dutch universities. This technique allowed us to extract structured latent variables from the program and course descriptions and to model the semantic relationships between topics and words.

The results indicate that the majority of the Dutch universities teach core skills such as research, ethics, and core terms, which likely form the foundation of data science and AI education. These findings are in line with the results of other studies that have emphasized the importance of ethics, research skills, and technical skills in data science education (Almgerbi et al., 2022, De Mauro et al., 2018; Fayyad & Hamutcu, 2020; Tang & Sae-Lim, 2016). Moreover, the results of our study reveal that while the prevalence of these topics is relatively consistent across universities, the order of the topics and the specific skills taught can vary significantly.

The results also suggest that the Dutch data science and AI programs are highly diverse, both in terms of the skillsets taught and the focus of the programs. For instance, the two technical universities, TU Delft and TU/e, score highly on electronics/IT, while the more general universities score higher on the research dimension. These findings are in line with the results of prior research that has observed a wide range in the skillsets taught across different data science programs (Tang & Sae-Lim, 2016; West, 2017). Our study provides a more nuanced view of this diversity, by showing that the diversity in the skillsets taught is not only dependent on the type of university, but also on the specific programs offered within each university.

The results of our study have important implications for various stakeholders. For universities and other educational institutions, our findings provide valuable insights into the content of data science and AI master programs and can be used to refine and update curricula. For businesses and organizations, the results can help to better match the qualifications of candidates to their specific needs. Finally, for current and aspiring data scientists, our findings can be used to identify gaps in their own skillsets and guide their professional development efforts.

However, it is important to note that our study has several limitations. Firstly, our results are based on the analysis of university program websites and may not reflect the actual skills taught in the classroom. Naturally, specific skillsets cannot be derived from the program texts on university website alone. As a result, it is hard to derive specific information from the text analyses without further knowledge of this field of science. Therefore, expert knowledge will be needed to fully understand what data science and artificial intelligence graduates are capable of doing. Additional attention will have to be put in by further research to get a good grasp on what the domains of data science and artificial intelligence exactly consist of.

Furthermore, both program and course descriptions have been used in these analyses. This leads to very broad descriptions and more narrowed down descriptions being put on the same pile and being considered as the same. In addition, some programs have a higher number of courses than other programs, which might cause the results to be skewed towards the programs with a higher number of courses (and thus a higher number of descriptions and therefore a higher number of words). These programs might have more influence on the results than programs with a lower number of courses or less detailed descriptions. This is also relevant for the posterior classification, since different universities have different totals of programs included in this study. Therefore, results might be skewed towards the universities with a higher number of programs included with more detailed descriptions.

Despite these shortcomings, the findings of the present study do shed light on the very ill defined concept of data science and artificial intelligence. By examining 1,009 master program descriptions, the most important aspects of data science and AI were uncovered, marking the beginning of a larger effort to fully define the knowledge domain of these graduates. This study fills a significant gap in literature by specifically focusing on the knowledge domain of these graduates and can be used to better define the emerging field from a practical perspective, as well as to bridge the gap between the supply of education and the demand of the job market. Further research could explore the skills that graduates actually use in their first jobs, and how this relates to their studies. Additionally, investigating the connection between the needs of specific market segments and the profiles of different universities could help to improve educational programs and assist students in making informed decisions about their future employment prospects.

Disclosure statement

# Appendix A

## Search keys Data Science and Artificial Intelligence

- knowledge map AND data science
- knowledge map AND data science OR AI OR artifical intelligence
- knowledge map AND artificial intelligence
- knowledge map AND computer science
- knowledge map AND data science OR AI OR artifical intelligence OR computer science
- definition of data science

# Appendix B

## Erasing non-informative words: "ad" and "hoc"

The words "ad" and "hoc got printed in one of topics of the LSA model. The words "Ad", "hoc" and "ad_hoc" would be stated here, which would contain 30% of the total shown words per topic, whereas the words would not hold any interpretable information. The words ad and hoc were only used in one description, but were used a lot and always together. This would lead to the LSA prioritizing this word to the extent where it would end up in one of the top 10 words of one of the 7 topics. This would lead to too much information being lost and therefore we decided to delete the words from the descriptions. Therefore the optimal number of K metrics and all analyses were rerun. Since the algorithms for finding an optimal number of K would run for two full days on a normal computer with a quad core processor. In light of not spending too much time we chose to ignore other combinations of words that would lead to some information being lost. An example of this case is in the LSA where k = 7 outcomes, topic 4 where "multimedia", "search" and "multimedia_search" would be stated. These cases erose after removing the words "ad" and "hoc" and therefore this might become an endless cycle of removing non-optimal words from the analyses.

# Appendix C

# Top ten most frequent words per topic for the CTM with 13 Topics

|   | Topic 1 | Topic 2 | Topic 3 | Topic 4 | Topic 5 | Topic 6 | Topic 7 |
|---|---|---|---|---|---|---|---|
| 1 | network | model | data | data | data | system | research |
| 2 | datascience | data | process | techniques | information | problem | design |
| 3 | programme | statistical | naturallanguageprocessing | algorithms | model | algorithms | scientific |
| 4 | data | analysis | techniques | analysis | programming | quantum | project |
| 5 | social | techniques | model | machinelearning | visualization | techniques | knowledge |
| 6 | research | methods | web | datamining | system | optimization | field |
| 7 | knowledge | theory | language | information | retrieval | artificialintelligence | researchproject |
| 8 | artificialintelligence | linear | datascience | knowledge | language | design | software |
| 9 | project | timeseries | information | theory | processing | methods | speech |
| 10 | information | regression | mining | applications | big | model | questions |
| Core | *Core terms* | *Statistical analyses* | *Natural language processing* | *Machine learning* | *Information* | *AI / optimization* | *Research* |

|   | Topic 8 | Topic 9 | Topic 10 | Topic 11 | Topic 12 | Topic 13 |
|---|---|---|---|---|---|---|
| 1 | image | system | deeplearning | business | project | model |
| 2 | machinelearning | design | reinforcement | data | research | system |
| 3 | model | security | algorithms | marketing | thesis | computational |
| 4 | data | data | model | datascience | economic | methods |
| 5 | techniques | health | neuronetwork | innovation | system | modeling |
| 6 | methods | business | system | management | analysis | human |
| 7 | analysis | management | methods | services | problem | processes |
| 8 | processing | software | deep | knowledge | data | theoretical |
| 9 | computer | digital | problem | concepts | services | neuroscience |
| 10 | vision | information | techniques | customer | start | cognitive |
| Core | Analyses | Information security / business | Machine learning | Business domain | Research | Modelling / Human brain |